\documentclass[3p,times,twocolumn]{article}

\usepackage{amssymb}
\usepackage{graphics}
\usepackage[tight,footnotesize]{subfigure}
\usepackage{multirow}
\usepackage{amsmath}
\usepackage{epstopdf}
\usepackage[figuresright]{rotating}
\usepackage{authblk}
\usepackage[acronym,nonumberlist]{glossaries}
\newacronym{ACU}{ACU}{Admission Control Unit}
\newacronym{CBR}{CBR}{Constant Bit Rate}
\newacronym{CFP}{CFP}{Contention Free Period}
\newacronym{CP}{CP}{Contention Period}
\newacronym{DCF}{DCF}{Distributed Coordination Function}
\newacronym{EDCA}{EDCA}{Enhanced Distributed Channel Access}
\newacronym{FTP}{FTP}{File Transfer Protocol}
\newacronym{HC}{HC}{Hybrid Coordinator}
\newacronym{HCCA}{HCCA}{HCF Controlled Channel Access}
\newacronym{HCF}{HCF}{Hybrid Coordination Function}
\newacronym{HTTP}{HTTP}{Hypertext Transfer Protocol}
\newacronym{MAC}{MAC}{Medium Access Control}
\newacronym{MSDU}{MSDU}{MAC Service Data Unit}
\newacronym{PCF}{PCF}{Point Coordination Function}
\newacronym{PHY}{PHY}{Physical Layer mode}
\newacronym{PIFS}{PIFS}{PCF Inter Frame Space}
\newacronym{QAP}{QAP}{QoS-enabled Acces Point}
\newacronym{QoS}{QoS}{Quality of Service}
\newacronym{QSTA}{QSTA}{QoS-enabled Station}
\newacronym{SMTP}{SMTP}{Simple Mail Transfer Protocol}
\newacronym{TBTT}{TBTT}{Target Beacon Transmission Time}
\newacronym{TXOP}{TXOP}{Transmission Opportunity}
\newacronym{TXOPs}{TXOPs}{Transmission Opportunities}
\newacronym{VBR}{VBR}{Variable Bit Rate}
\newacronym{TGe}{TGe}{IEEE 802.11 Task Group E}
\newacronym{WM}{WM}{Wireless Medium}
\newacronym{CAP}{CAP}{Controlled Access Phase}
\newacronym{AC}{AC}{Access Categorie}
\newacronym{UGC}{UGC}{User-Generated Content}
\newacronym{TS}{TS}{Traffic Stream}
\newacronym{TSPEC}{TSPEC}{TS Specification}
\newacronym{SI}{SI}{Service Interval}
\newacronym{BSS}{BSS}{The Basic Service Set}
\newacronym{QS}{QS}{Queue Size}
\newacronym{IFS}{IFS}{Interframe Space}
\newacronym{HDTV}{HDTV}{High Definition TV}
\makeglossaries
\begin{document}
	\title{Providing Dynamic TXOP for QoS Support of Video Transmission in IEEE 802.11e WLANs}
	
\author[1]{Mohammed A. Al-Maqri}
\author[1]{Mohamed Othman}
\author[2]{Borhanuddin Mohd Ali}
\author[1]{Zurina Mohd Hanapi}	

\affil[1]{Department of Communication Technology and Network}
\affil[2]{Department of Computer and Communication Systems Engineering, Universiti Putra Malaysia, 43400 UPM, Malaysia}

\maketitle
\begin{abstract}
The IEEE 802.11e standard introduced by \gls{TGe} enhances the \gls{QoS} by means of  \gls{HCCA}. The scheduler of \gls{HCCA} allocates \glspl{TXOP} to \gls{QSTA}  based on their \glspl{TSPEC} negotiated at the traffic setup time so that it is only efficient for \gls{CBR} applications. However, \gls{VBR} traffics are not efficiently supported as they exhibit non-deterministic profile during the time. In this paper, we present a dynamic \gls{TXOP} assignment Scheduling Algorithm for supporting the video traffics transmission over IEEE 802.11e wireless networks. This algorithm uses a piggybacked information about the size of the subsequent video frames of the uplink traffic to assist the Hybrid Coordinator accurately assign the \gls{TXOP} according to the fast changes in the \gls{VBR} profile. The proposed scheduling algorithm has been evaluated using simulation with different variability level video streams. The simulation results show that the proposed algorithm reduces the delay experienced by \gls{VBR} traffic streams comparable to \gls{HCCA} scheduler due to the accurate assignment of the \gls{TXOP} which preserve the channel time for transmission.
\end{abstract}	
\section{Introduction}
\label{sec:intro}
Due to the wide spread of ubiquitous applications in the internet and the rapid growth of multimedia streams, providing differentiated \gls{QoS} for such applications in Wireless Local Area Networks (WLANs) has become a very challenging task. The IEEE802.11 \cite{IEEEStand1999} has become the most deployed technology in WLANs due to some of its key features like deployment flexibility, infrastructure simplicity and cost effectiveness. IEEE802.11 introduces two channel access modes, namely \gls{DCF} and \gls{PCF}. The former is the mandatory medium access method which is appropriate to serve best effort applications such as \gls{HTTP}, \gls{FTP} and \gls{SMTP}. Multimedia streams that require a certain \gls{QoS} level are served during the controlled mode (i.e. PCF) since it provides a contention-free polling-based access to the channel to provide the demanded \gls{QoS}. However, it is not efficient enough to support high \gls{QoS} requirement applications due to the fact that \gls{PCF} only operates on the Free-Contention period, which may noticeably cause an increase in the transmission delay especially with high bursty traffics. Consequently, IEEE 802.11 \gls{TGe} has presented IEEE 802.11e protocol \cite{IEEEStandard2007} and revised version \cite{IEEEStandard2012} with new technical enhancements on \gls{MAC} and Physical layer.

IEEE 802.11e introduces \gls{HCF} which extends the \gls{MAC} of IEEE 802.11 standard. \gls{EDCA} function is an extension to \gls{DCF}, which operates in a distributed manner to provide prioritized \gls{QoS}. \gls{HCCA} is an extension of \gls{PCF} that introduces a polling mechanism to provide parameterized \gls{QoS} for applications that require rigorous \gls{QoS} requirements. \gls{EDCA} introduces a random access to the wireless medium by means of access categories (ACs). The traffics are mapped to ACs according to their priority. Every \gls{AC} will be associated with a backoff timer so that the highest priority ACs will go through a shorter backoff process. Despite \gls{EDCA} provides \gls{QoS} support, it is still not efficient for application with rigid \gls{QoS} requirements. Delay-sensitive multimedia streams are more adequate to be transmitted throughout \gls{HCCA} since it was designated to minimize the overhead of messaging caused by the distributed approach of \gls{EDCA} and thus guarantee the required better \gls{QoS}. In \gls{HCCA}, the \gls{HC} polls wireless stations periodically and allocates \gls{TXOP} to them. And yet, \gls{HCCA} schedules traffics upon their \gls{QoS} requirements negotiated in the first place, it is only suitable for \gls{CBR} applications such as G.711 \cite{G7111988}, audio streams, and (H.261/MPEG-1) video \cite{MPEG11997}. The allocation of the \gls{TXOP} based on the mean characteristics negotiated at the traffic setup is not accurate, because of the deviation of \gls{VBR} traffics from its mean characteristics. By 2014, about 91\% of web traffic will be video streams \cite{DigitalMedia2012}. For this reason several researches have been  carried out to improve the performance of WLANs in terms of provisioning \gls{QoS} for such streams. Motion Picture Experts Group type 4 (MPEG--4/H.264) has become a prominent video the internet due to its scalability, error robustness and network-friendly features. \gls{HCCA} is not convenient to deal with the fluctuation of the \gls{VBR} traffic such as MPEG--4 streams, where the packet size shows high variability during the time. This consequently leads to a remarkable increase in the end-to-end delay of the delivered traffics and degradation in the channel utilization.

With the increase of Internet web applications in the wireless mobile devices, the \gls{UGC} such as pre-recorded video streams has become more prominent nowadays. To the best of our knowledge, the scheduling of uplink pre-recorded continuous media in \gls{HCCA} has not been addressed efficiently despite the fast growth of uplink streams of the \gls{UGC} in the Internet such as pre-recorded video streams. In this paper, we present an enhancement on the \gls{HCCA} scheduling algorithm aiming to adapt to the fast fluctuation of \gls{VBR} video traffics profile. Basically, the proposed scheduling algorithm makes use of the fact that in the video applications that use prerecorded streams, the video traffic can be analyzed prior to the call setup. This fact has been highlighted in Feed Forward Bandwidth Indication \ .It computes the \gls{TXOP} for a traffic based on knowledge about the actual frame size instead of assigning \gls{TXOP} according to mean characteristics of the traffic which is unable to reflect the actual traffic. This algorithm uses the \gls{QS} of IEEE 802.11e \gls{MAC} header to carry this information to the \gls{HC}.

The rest of the paper is organized as follows. Section ~\ref{sec:relatedWorks} illustrates the reference \gls{HCCA} mechanism and its deficiency in supporting \gls{VBR} Video streams and demonstrates some of the \gls{HCCA} related works. Section ~\ref{sec:ATAV} illustrates the proposed dynamic assignment algorithm. The performance evaluation and results discussion is shown in Section ~\ref{sec:evaluation}. Section ~\ref{sec:conclusion} concludes the work presented in this paper.

\section{Background and Related Works}
\label{sec:relatedWorks}
This section describes IEEE 802.11e \gls{HCCA} scheduler and some characteristics of MPEG--4 VBR video traffic. The deficiency of \gls{HCCA} in supporting \gls{VBR} is illustrated and some related works in enhancing its performance are also discussed.

\subsection{IEEE 802.11e HCCA Mechanism}
\begin{figure*}
	\centering
	\includegraphics[width=\linewidth]{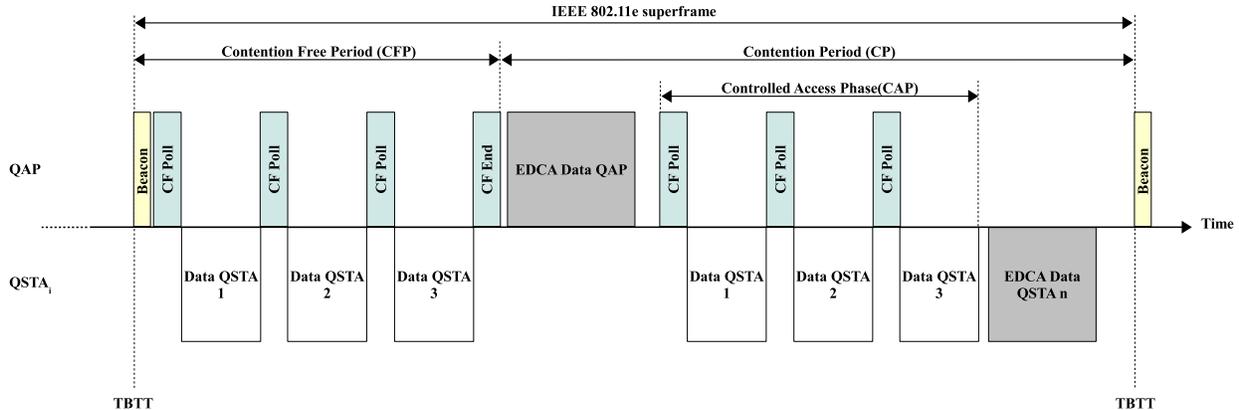}
	\caption{Controlled channel access mechanism in IEEE 802.11e HCCA.}
	\label{fig001}
\end{figure*}

In IEEE 802.11e, a parameterized \gls{QoS} is supported during \gls{HCCA} using polling access method. A beacon is transmitted every \gls{TBTT} comprising a superframe which in turn includes \gls{CFP} followed by \gls{CP}. The \gls{HC} shall initiate a \gls{CFP}, to deliver its data traffics, or allocate a \gls{TXOP} to a \gls{QSTA} in \gls{CP} to allow uplink traffics to be transmitted. In both cases the \gls{HC} senses the \gls{WM}. When the \gls{WM} is found idle for a \gls{PIFS} period, the \gls{HC} shall transmit its data during \gls{CFP} period or permit a \gls{QSTA} to start a frame exchange sequence with \gls{HC} to cover the allocated \gls{TXOP} duration. The \gls{HC} may begin a \gls{CAP} at any time during the \gls{CP} if the medium remains idle for a time equals to \gls{PCF} \gls{PIFS}.  \gls{HCCA} outperforms \gls{PCF} of legacy IEEE802.11, in that it can be initiated in both \gls{CFP} and \gls{CP} in contrary to its ancestor, \gls{PCF}, which only operates during  \gls{CFP}. When a station intends to initiate a data traffic, it issues a \gls{QoS} reservation through a special \gls{QoS} management action frame called ADDTS-Request contains a set of parameters that define the characteristics of the \gls{TS} (TSPEC). The fields of the \gls{TSPEC} and how the \gls{HC} exploits them in the scheduling process is discussed in details in the next section. Figure~\ref{fig001} demonstrates an example of \gls{HCCA} transmission during \gls{CFP} and \gls{CP} periods.

\subsection{HCCA Scheduler}
\label{sec:HCCA}
As mentioned earlier, in order to initiate an uplink traffic, the QSTA issues a \gls{QoS} reservation through transmitting an ADDTS-Request frame. This frame carries information about the TSPEC which is required by HC for scheduling purpose. The mandatory fields of the TSPEC are described as follows:
\begin{itemize}
	\item Mean Data Rate($\rho$): average data packet rate measured in units of (bits / seconds).
	\item Nominal MSDU length($L$): mean size of MAC packets in units of bytes.
	\item Maximum MSDU Size($M$): maximum allowable size of the MAC packet in the \textit{TS} in units of bytes.
	\item Delay Bound($D$): maximum allowable delay for a packet to be transmitted through the wireless medium in units of milliseconds.
	\item Service Interval($SI$): time period between a station's \gls{TXOPs} measured in units of milliseconds.
	\item Physical Rate($R$): the assumed wireless physical bit rate, measured in of (bits/second).
\end{itemize}

The HC which usually resides in the \gls{QAP} maintains the \glspl{TSPEC} of all \textit{TSs} in the so-called polling list. Accordingly, \gls{HC} computes the duration of the time to be granted to each \gls{QSTA} for the transmission of their traffics (TXOP). The admission of the \textit{TSs} is governed by \gls{HC}, using the \gls{ACU}. \gls{HC} reserves the right to accept or reject any \textit{TS} so as to preserve the \gls{QoS} of the previously admitted \textit{TSs}. If \gls{HC} accepts the traffic it will respond by an ADDTS-Response or a rejection message otherwise.

Upon receiving an ADDTS-Request from a \gls{QSTA}, the \gls{HCCA} scheduler of the \gls{HC} goes through the following steps to schedule the uplink traffics:
\begin{enumerate}
	\item \textit{SI Assignment}
	
	\label{eq:SIassign}
	The scheduler calculates $SI$ as the minimum of all Maximum Service Intervals ($MSI$) of all admitted traffic streams which is a submultiple of the beacon interval. The minimum $MSI$ for each \gls{QSTA} is obtained from Equation~\eqref{eq:minSI}.
	\begin{equation}
		\label{eq:minSI}
		MSI_{min}= min(MSI_{i}),\quad i=1,2,3,\ldots,n
	\end{equation}
	where $n$ is the number of admitted \textit{TSs} and $MSI_{i}$ is the maximum $SI$ of the $i^{th}$ stream.
	The $SI$ is computed so that it satisfies the condition in Equation~\eqref{eq:si}.
	\begin{equation}
		\label{eq:si}
		SI=\frac{BeaconInterval}{x}\leq MSI_{min},
	\end{equation}
	where the denominator, $x$, is an integer number that divides the beacon interval into the largest number that is equal or less than the $MSI_{min}$.\\
	\item \textit{TXOP Allocation}
	
	\gls{HC} allocates a \gls{TXOP} to each admitted \gls{QSTA} so as to enable it to transmit its data with regards to the negotiated \gls{QoS} parameters of the \gls{TSPEC}.
	
	For the $i^{th}$ \gls{QSTA}, the scheduler computes the number of MSDUs arrived at $\rho_{i}$ as in Equation~\eqref{eq:n}.
	\begin{equation}
		N_{i}=\left \lceil \frac{SI\times\rho_{i}}{L_{i}} \right \rceil,
		\label{eq:n}
	\end{equation}
	where $L_{i}$ is the nominal \gls{MSDU} length for the $i^{th}$ \gls{QSTA}.
	Then the \gls{TXOP} duration of the particular station, $TXOP_{i}$, is calculated as the maximum of the time required to transmit $N_{i}$ \gls{MSDU} or the time to transmit one maximum \gls{MSDU} at the physical rate $R_{i}$, as stated in Equation~\eqref{eq:txop}.
	\begin{equation}
		TXOP_{i}=max\left (\frac{N_{i}\times L_{i}}{R_{i}}+O,\frac{M}{R_{i}}+O \right)
		\label{eq:txop}
	\end{equation}
	Where $M$ is the maximum \gls{MSDU} and $O$ denotes the overhead, including \gls{MAC} and \gls{PHY} headers, \glspl{IFS}, and the acknowledgment and poll frames overheads.\\
	\item \textit{Admission Control}
	
	The \gls{ACU} manages the \textit{TSs} admission while maintaining the \gls{QoS} of the already admitted ones. When a new \textit{TS} demands an admission, the \gls{ACU} First obtains a new $SI$ as shown in the previous step and computes number of MSDUs arrived at the new $SI$ using Equation~\eqref{eq:n}. Next, it calculates the $TXOP_{i}$ for the particular \textit{TS} using Equation~\eqref{eq:txop}. Finally, \gls{ACU} admits only the \textit{TS} if the following inequality is satisfied.
	\begin{equation}
		\label{eq:ACU}
		\frac{TXOP_{n+1}}{SI}+\sum_{i=1}^{n} \frac{TXOP_{i}}{SI}\leq\frac{T- T_{CP}}{T}
	\end{equation}
	where $n$ is the number of currently admitted \textit{TSs}, so that ($n+1$) represents the incoming \textit{TS}, $T$ is the time interval between two consecutive TBTTs periods, beacon interval and $T_{CP}$ is the duration reserved for \gls{EDCA}. The \gls{HC} sends an acceptance message (ADDTS-Response) to the requested \gls{QSTA} if the condition in Equation~\eqref{eq:ACU} is true or send a rejection message otherwise. The accepted \textit{TS} will be added to the polling list of the \gls{HC}.
\end{enumerate}
\subsection{Variable Bit Rate MPEG--4 Video Traffic}
\label{sec:VBRTraffic}
MPEG--4 is an efficient video encoding covering a wide domain of bit rate coding ranging from low-bit-rate for wireless transmission up to higher quality beyond \gls{HDTV} \cite{Fitzek2001}. For this reason, MPEG--4 video coding has become from among the prominent videos in the internet nowadays.

In fact, MPEG--4 videos are encoded using different compression ratios which produce different levels of quality. Higher compression level generates lower-quality video with smaller mean frame sizes and smaller mean bit rate and vice versa. This variability in the compression level is adequate to transmit the video packets over the limited wireless network resources such as low bit rate. Table~\ref{tab:traceFragHigh} displays excerpt of video trace file of Jurassic Park 1 movie \cite{Fitzek2000} encoded using MPEG--4 at high quality.
In MPEG--4 video coding, successive pictures of the coded video stream compose a Group of Picture (GoP) which identifies how the intra- (I-frame) and inter-frames (P- and B- frames) are ordered, we refer the reader to \cite{soares1998,koenen1999,Koenen2002} for more details about MPEG--4 videos.
In this excerpt, we display one GoP of encoded video which consists the pattern IBBPBBPBBPBB. One can notice that the frame are not sequenced chronologically, yet it is ordered according to the display time instead.
\begin{table*}
	\centering
	\caption {A fragment of Jurassic Park 1 Trace File Encoded Using MPEG--4 at High Bit Rate}
	\begin{tabular}{cccc}
		\hline
		Frame sequence & Frame type & Frame period (ms) & Frame size (byte)
		\\ \hline
		527 & I & 21120 & 8124 \\
		528 & B & 21040 & 6442 \\
		529 & B & 21080 & 6237 \\
		530 & P & 21240 & 7581 \\
		531 & B & 21160 & 6184 \\
		532 & B & 21200 & 6173 \\
		533 & P & 21360 & 7482 \\
		534 & B & 21280 & 6331 \\
		535 & B & 21320 & 6567 \\
		536 & P & 21480 & 7130 \\
		537 & B & 21400 & 6410 \\
		538 & B & 21440 & 6223 \\ \hline
	\end{tabular}
	\label{tab:traceFragHigh}
\end{table*}
As it is mentioned above, \gls{HC} schedules \glspl{QSTA} with respect to their negotiated \gls{TSPEC} parameters which represent the mean characteristics of the traffics. Basically, the weakness of \gls{HCCA} in supporting \gls{VBR} traffic is because of the lack of information about the abrupt changes of the traffic profile during the time, more particular the traffic burstiness issue. In the case of the transmission of prerecorded uplink	 video traffics, it will be beneficial to inform the \gls{HC} about the changing in the video profile to accommodate the fast fluctuation of the traffic.

\hspace{1mm}Several approaches such as \cite{Lee2009,Jansang2011,Cecchetti2012,cecchettielAL2012,ruscelli2013} have been presented in the literature attempting to remedy the deficiency of the \gls{HCCA} reference scheduler in supporting \gls{QoS} for \gls{VBR} traffics. However, these enhancements are still not sufficient to cope with the fast fluctuating nature of highly compressed video applications since the \glspl{QSTA} are scheduled according to an estimation about the uplink \textit{TSs} characteristic which may be far from the real traffics.
\subsection{Transmission of MPEG--4 video in HCCA}
\begin{figure*}[!t]
	\centering
	\includegraphics[width=\linewidth]{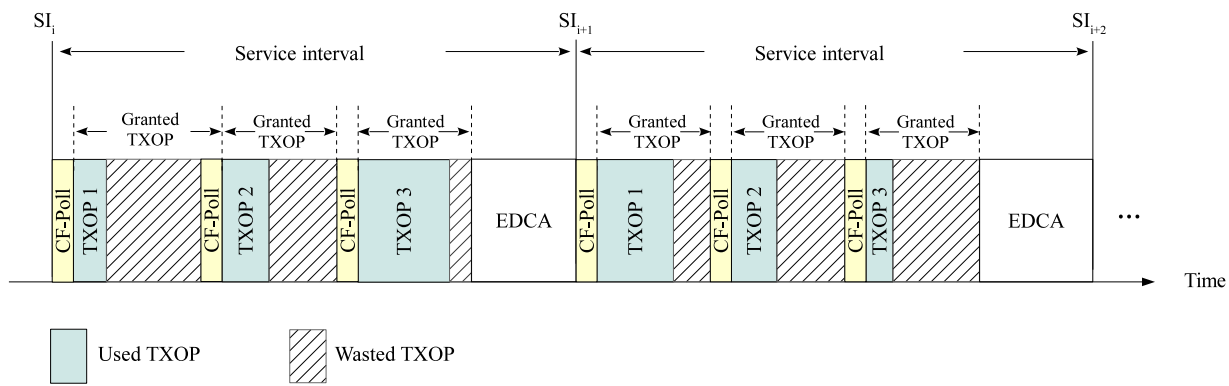}
	\caption{Example of MPEG--4 video transmission using HCCA.}
	\label{dig:wastedTXOP}
\end{figure*}

Although \gls{HCCA} guarantees a \gls{QoS} for video traffic based on the required \gls{TSPEC} parameters, there is a probability to have frames smaller than the mean negotiated \gls{MSDU} size. Consequently, a larger \gls{TXOP} than needed will be assigned to a \gls{QSTA} causing wasting in wireless channel time and remarkable increase in the end-to-end delay. Figure~\ref{dig:wastedTXOP} illustrates the effect of assigning \gls{TXOPs} for \gls{VBR} traffics based on the mean \gls{TSPEC} parameters on increasing the packet delay and on the poor wireless channel utilization. Suppose there are three stations sending uplink video traffics to \gls{QAP}. \gls{HC} will accordingly assign $TXOP_{1}$, $TXOP_{2}$ and $TXOP_{3}$  to $QSTA_{1}$, $QSTA_{2}$ and $QSTA_{3}$ respectively. Assume that in any $SI$ some or all the frames sent is considerably smaller than the negotiated \gls{MSDU} for the \textit{TS}, in this case the \gls{QSTA} will only utilize a portion of the scheduled \gls{TXOP} for sending its data as explained in the $SI_ {I} $ and $SI_{i+1}$. The next scheduled \gls{TXOP} will initiate according to its scheduled time regardless the actual exploited time in the previous \gls{TXOP} causing increases in the delay and wasting the channel time as well. This issue may be noticeably severe when the number of \gls{QSTA} with \gls{VBR} traffics increase. Moreover, in the transmission of the prerecorded video, the traffic behavior is known prior to the traffic setup. These observations motivate us to present an enhancement to \gls{HCCA} scheduler in which \gls{HC} exploits information sent by \gls{QSTA} about the changes in the traffic profile so as to accurately assign \gls{TXOP} to \gls{QSTA} and advance the consecutive \gls{TXOPs} to minimize the delay and add the residual wireless channel time to \gls{EDCA} period. The proposed scheduling algorithm is presented in details in the next section.
\section{Dynamic TXOP Assignment Scheduling Algorithm}
\label{sec:ATAV}
\begin{figure*}[!t]
	\centering
	\includegraphics[width=\linewidth]{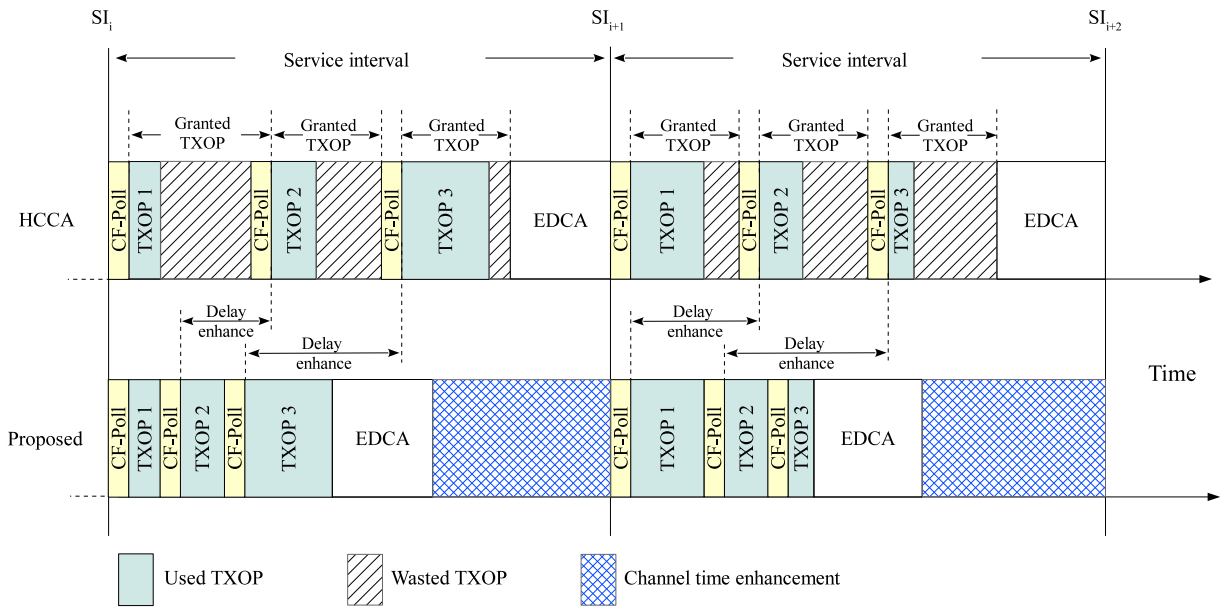}
	\caption{Dynamic TXOP assignment scheduling algorithm.}
	\label{dig:HCCAvsATAV}
\end{figure*}

\gls{HCCA} scheduler computes the \gls{TXOP} duration by estimating the amount of data expected to be transmitted by the \gls{QSTA} during \gls{SI}. This estimation is based on the \gls{TSPEC} negotiated with \gls{HC} which considers the mean characteristics of the traffic, Equation~\eqref{eq:txop}. The proposed scheduling algorithm described in this section is referred to as Dynamic Transmission Opportunity because it adapts \gls{TXOP} duration based on the feedback of the next frame packet size reported by \glspl{QSTA}. The proposed algorithm gives an actual \gls{TXOP} needed by stations and ensures that the delay is minimized without jeopardizing the channel bandwidth. The delay experienced by the proceeding unused(wasted) \gls{TXOPs} is minimized using this algorithm as illustrated in Figure~\ref{dig:HCCAvsATAV}. The scheduling parameters along with the scheduling operation are described below.
\subsection{Scheduling Parameters}
As the proposed scheduler operates based on the feedback, the \gls{HCCA} scheduler will change some scheduling parameters in Equation~\eqref{eq:txop} upon receiving a feedback from \gls{QSTA}. Herein a description of these parameters:

\subsubsection{Number of MSDUs Received in \gls{SI} ($N_{i}$)} Using Equation~\eqref{eq:n}, the reference scheduler calculates the expected number of the received packets every \gls{SI} based on the mean \gls{TSPEC} parameters at the traffic setup phase. In our algorithm, this parameter is set to 1 as only one packet is expected to be received every \gls{SI}.
\label{enm:N_i}
\subsubsection{Mean Size of MSDU ($L_{i}$)} The \gls{HC} updates $L_{i}$ in Equation~\eqref{eq:txop} with regards to the information piggybacked with each packet received from a \gls{QSTA}. In fact, this is the major part of the proposed algorithm in which the \gls{TXOP} duration given to a \gls{QSTA} is calculated dynamically to accommodate the actual packet size to be received at the \gls{QAP} from an uplink \textit{TS}.
\label{enm:L_i}
\subsection{The Mechanism of Dynamic TXOP assignment}
In this algorithm, the exact \gls{MSDU} size of the next frame of the uplink stream is obtained from the application layer through cross layering. This information is transmitted with each packet to the \gls{QAP} carrying the next frame size. Upon a data frame reception, the \gls{HC} recalculates the \gls{TXOP} duration to be granted for a particular station in the next \gls{SI} so as to adapt to the fast varying in \gls{VBR} video traffic. Consequently, it minimizes the packet end-to-end delay and conserve the channel bandwidth. In this section, we present the description of \gls{TXOP} operation at both \glspl{QSTA} and the \gls{QAP}.
\subsubsection{Operation at the station}
At the \gls{QSTA}, information about the next \gls{MSDU} frame size is obtained from the application layer via cross-layering. This information is carried in the \gls{QS} field introduced by IEEE 802.11 standard \cite{IEEEStandard2012} which is a part of the \gls{QoS} Control field of the \gls{QoS} data frame. The \gls{QS} field is exploited in this approach for sending information about the next \gls{MSDU} frame size to the \gls{QAP}.
\subsubsection{Operation at the access point}
\begin{figure*}[!t]
	\centering
	\includegraphics[width=\linewidth]{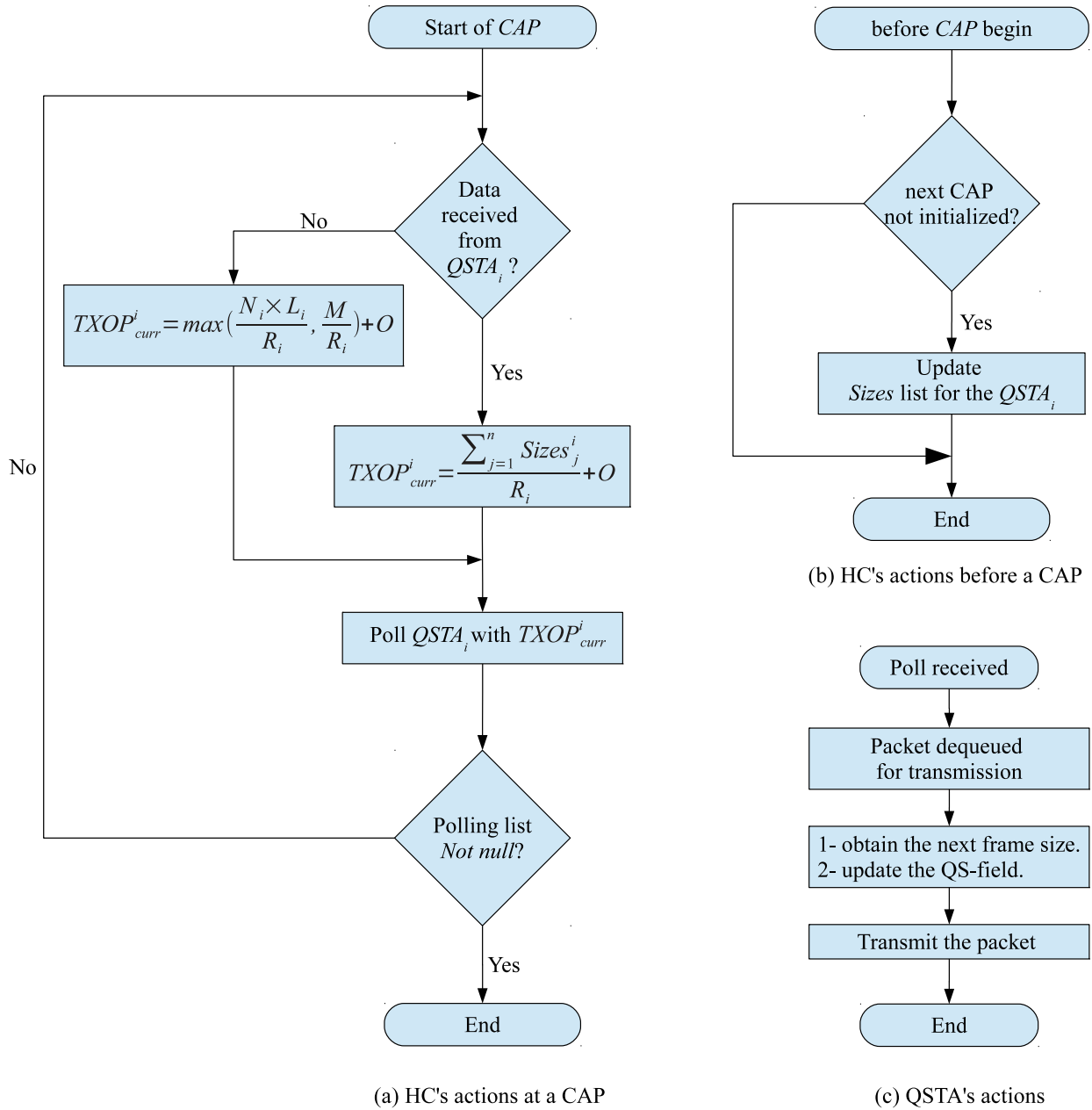}
	\caption{Flow chart of the proposed scheduling algorithm}
	\label{flowchart}
\end{figure*}

After the traffic setup phase, the \gls{QAP} transmits the first poll frame granting the \gls{QSTA} a \gls{TXOP} duration. The station will accordingly transmit the first packet of its traffic to the \gls{QAP}. Note that the inter-arrival time between encoded video traffic frames is a multiple of a fixed interval (typically 40 ms) depends on the encoding parameters. That is to say, it is expected to receive only one packet at a multiple of a designated interval. Details about the operation of our approach at \gls{QAP} is reported in Fig.~\ref{flowchart}.


At the beginning of each \gls{CAP}, \gls{HC} goes through the $stations_{i}$ list and computes $TXOP_{i}$ for the $station_{i}$ according to one the these cases: first case is when a data packet is received from the $station_{i}$ in the previous CAP/SI period, the \gls{MSDU} size ($Size_{i}$) of the next frame is obtained from the \gls{QS} field of IEEE 802.11e \gls{MAC} header. Then, a $TXOP_{i}$ of $QSTA_{i}$ is calculated using Equation~\eqref{eq:atav}. In other words, the \gls{TXOP} in $SI_{i+1}$ is scheduled based on the information received by QAP/HC during $SI_{i}$, as depicted in Figure~\ref{dig:HCCAvsATAV}.
\begin{equation}
	TXOP_{i}=\frac{Size_{i}}{R_{i}}+O
	\label{eq:atav}
\end{equation}
The other case is when no data packet is received due to loss, the \gls{QAP} will use the Equation~\eqref{eq:txop} of  \gls{HCCA} scheduler to compute the $TXOP_{i}$. It is worth noting that at the first \gls{CAP} of any \textit{TS}, the \gls{TXOP} is calculated based on Equation~\eqref{eq:txop} because no information about the next packet size has been reported yet.

\section{Performance Evaluation}
\label{sec:evaluation}
In order to evaluate the performance of the proposed scheduler, we have used a network simulation tool. The simulation environment setup, and video traffic used as uplink traffics is described in details in this section. The performance of our scheduler is compared against the \gls{HCCA}. The results of end-to-end delay and throughput are also discussed.
\subsection{Simulation Setup}
The proposed scheduler has been implemented in the well-known network simulator (\textit{ns-2}) \cite{NS2} version 2.27. The \gls{HCCA} implementation framework \textit{ns2hcca} \cite{cicconetti2005} has been patched to provide the controlled access mode of IEEE 802.11e functions, \gls{HCCA}. The \textit{ns-2} Traffic Trace \cite{NSBook2012} agent is used for video stream generation.

\begin{figure}
	\centering
	\includegraphics[scale=0.43]{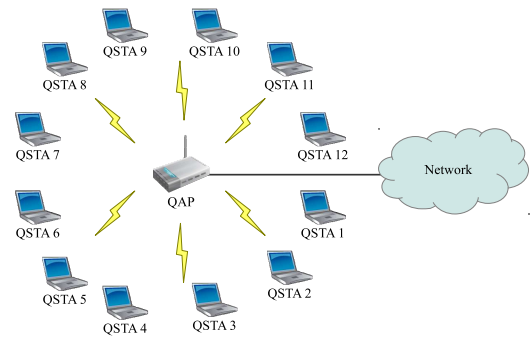}
	\caption{Network topology.}
	\label{fig:topology}
\end{figure}

A star topology has been used for constructing the simulation scenario to form an infrastructure network with one \gls{QAP} surrounded by varying number of the \glspl{QSTA} ranging from 1 to 12. All \glspl{QSTA} were distributed uniformly around the \gls{QAP} with a radius of 10 meters as shown in Figure~\ref{fig:topology}. The stations were placed within the \gls{QAP} coverage area, in the same basic service set \gls{BSS}, and the wireless channel is assumed to be ideal. Since we focus on \gls{HCCA} performance measurement, all the stations operate only on the contention-free mode by setting $T_{CP}$ in Equation~\eqref{eq:ACU} to zero.
\gls{QAP} is the sink receiver, while all stations are the video sources. Each send only an uplink video traffic as only one flow per station is supported in \textit{ns2hcca} patch. Therefore, for simulating concurrent video streams multiple stations are added each with one flow. In order to leave an ample time for initialization, stations start their transmission after 20 (sec) from the start of the simulation time and last until the simulation end. Wireless channel assumed to be an error-free, and no admission control used for the sake of investigating the maximum scheduling capability of each examined algorithm under heavy traffic conditions. Simulation parameters are summarized in Table~\ref{tab:SimPars}.
\begin{table}
	\caption {Simulation Parameters}
	\centering
	
	\begin{tabular}{ll}
		\hline
		Parameter				& Value			\\ \hline
		Simulation time			& 500 (sec)		\\
		Physical layer			& IEEE 802.11b	\\
		MAC layer				& IEEE 802.11e	\\
		SIFS					& 10 $\mu s$	\\
		PIFS					& 30 $\mu s$	\\
		Slot time				& 20 $\mu s$	\\
		Physical preamble length& 18 (bytes)	\\
		PLCP header length		& 6 (bytes)		\\
		PLCP data rate			& 1 (Mbps)		\\
		MAC header size			& 36 (bytes)	\\
		Data rate				& 11 (Mbps)		\\
		Basic physical rate		& 1 (Mbps)		\\ \hline
	\end{tabular}
	\label{tab:SimPars}
\end{table}
For evaluating the performance of our scheduling algorithm against the reference scheduler of \gls{HCCA}, Jurassic Park 1 video sequence trace encoded using MPEG--4 was chosen from a publicly available library for video traces \cite{Fitzek2001}. We tested the proposed scheduler on \textit{Jurassic Park 1} and \textit{Formula 1 } trace files which can be classified into movie and sport, respectively, which show different variability level. Table~\ref{tab:traceStats} demonstrates some statistics of the examined traces. The selected video is encoded using different Compression ratio, which results in varying quality.
\begin{table*}
	\centering	
	\caption {  Frame Statistics of MPEG--4 Video Trace Files}
	\begin{tabular}{ l|p{5cm}llll }
		\hline
		& & \multicolumn{2}{c}{Video Quality} \\
		\cline{3-4}
		Video & Parameter 			& Low quality	& High quality	\\ \hline
		\multirow{6}{*}{Jurassic Park 1}& Comp. ratio (YUV:MP4)		& 49.46			& 9.92			\\
		& Mean size (byte)			& 7.7e+02		& 3.8e+03		\\
		& CoV of frame size			& 1.39			& 0.59			\\
		& Mean bit rate (bit/sec)	& 1.5e+05		& 7.7e+05		\\
		& Peak bit rate (bit/sec)	& 1.6e+06		& 3.3e+06		\\
		& Peak/Mean of bit rate		& 10.61			& 4.37			\\ \hline
		\multirow{6}{*}{Formula 1}		& Comp. ratio (YUV:MP4)		& 43.51			& 9.92			\\
		& Mean size (byte)			& 8.7e+02		& 4.2e+03		\\
		& CoV of frame size			& 1.12			& 0.42			\\
		& Mean bit rate (bit/sec)	& 1.7e+05		& 8.4e+05		\\
		& Peak bit rate (bit/sec)	& 1.4e+06		& 2.9e+06		\\
		& Peak/Mean of bit rate		& 8.05			& 3.45			\\ \hline
	\end{tabular}
	\label{tab:traceStats}
\end{table*}
In this paper, we have tested the schedulers with low and high quality video trace. It is worth noting that the variability of the selected videos is measured by the Coefficient of Variation (CoV), which is the standard deviation of the frame size divided by the average frame size. TSPEC parameters used for each video traffic is shown in Table~\ref{tab:VideoParas} with regards to video \gls{QoS} requirements.
\begin{table}[!h]
	\centering	
	\caption {Video Traffic Parameters}
	\begin{tabular}{ l|lllll }
		\hline
		& & \multicolumn{2}{c}{Video Quality} \\
		\cline{3-4}
		Video	& Parameter	& Low quality	& High quality	\\ \hline
		\multirow{6}{*}{\rotatebox[]{90}{Jurassic Park 1}}& $L$	(bytes)		& 7.7e+02		& 3.8e+03 		\\
		& $M$ (bytes)		& 8154			& 16745 		\\
		& $\rho$ (bit/sec)	& 1.5e+05		& 7.7e+05		\\
		& $D$ (sec)		& 0.08			& 0.08			\\
		& $R$ (Mbps)		& 11			& 11			\\
		& $MSI$	(sec)		& 0.04			& 0.04			\\ \hline
		\multirow{6}{*}{\rotatebox[]{90}{Formula 1}}		& $L$ (bytes)		& 8.7e+02		& 4.2e+03 		\\
		& $M$	(bytes)		& 7032			& 14431 		\\
		& $\rho$ (bit/sec)	& 1.7e+05		& 8.4e+05		\\
		& $D$ (sec)		& 0.08			& 0.08			\\
		& $R$ (Mbps)		& 11			& 11			\\
		& $MSI$	(sec)		& 0.04			& 0.04			\\ \hline
	\end{tabular}
	\label{tab:VideoParas}
\end{table}
\subsection{Results and Discussion}
Simulations have been carried out to exhibit the performance of the examined schedulers using a different variability level of the same videos. Since the main objective is to achieve superior \gls{QoS} support by accurately granting \gls{TXOP} to the station to fit its need, packet end-to-end delay of the uplink traffics has been measured which considered as one of the significant metrics to evaluate a \gls{QoS} support of video streams. To validate the behavior of the examined schedulers, the measurements are done for an increasing number of \textit{TSs}. The system throughput was also investigated to verify that the improvement in the delay is achieved without jeopardizing the channel bandwidth.

\begin{figure*}
	\centering
	\subfigure[Low-quality.]{
		\label{fig:AllocatedTXOPLow1}
		\includegraphics[scale=0.70]{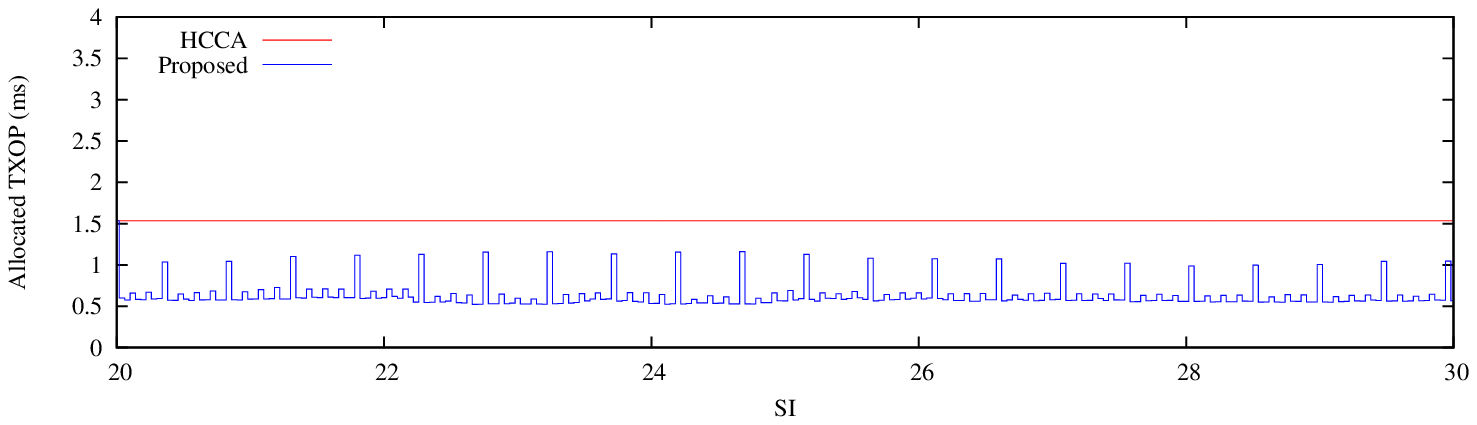}
	}
	\subfigure[High-quality.]{
		\label{fig:AllocatedTXOPHigh1}
		\includegraphics[scale=0.70]{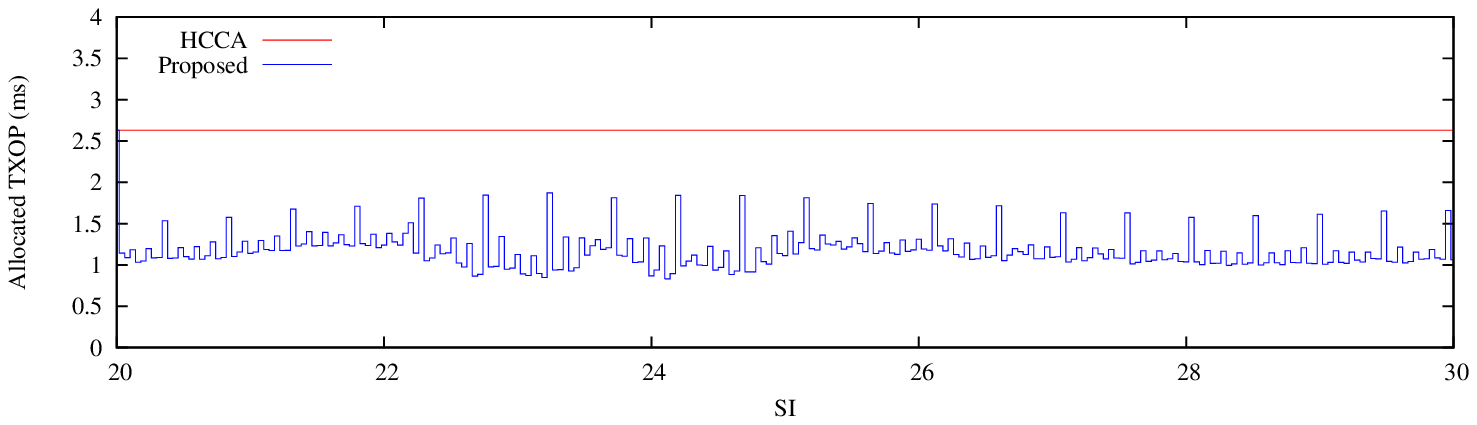}
	}
	\caption{TXOP allocation of Formula 1 video}
	\label{fig:allocTXOP}
\end{figure*}

The behavior of the examined schedulers in terms of allocating \gls{TXOP} in each \gls{SI} is illustrated in  Figure \ref{fig:allocTXOP} for the Formula 1 video sequence. The allocated \gls{TXOP} for one flow is shown against number of SIs for a duration of 10 seconds. The results reveal the fact of assigning fixed \gls{TXOP} in \gls{HCCA} for all SIs of the flow with accordance to Equation~\eqref{eq:txop}. In this case, the \gls{HCCA} computes \gls{TXOP} duration based on the maximum \gls{MSDU} size of the flow, namely 7032 bytes and 14431 bytes for low and high quality video respectively. Nevertheless, the proposed scheduler adaptively allocates a \gls{TXOPs} for each \gls{SI} based on the actual frame size obtained from the feedback information which show that in some SIs the allocated \gls{TXOP} duration in \gls{HCCA} is much higher than the actual need of the flow which considered as over-allocation cases. It is obvious that the \gls{TXOP} duration given in \ref{fig:AllocatedTXOPHigh1} is higher than that in \ref{fig:AllocatedTXOPLow1} as the mean bit rate of high quality encoded Formula 1 is considerably higher than that in low quality video sequence, refer to Table \ref{tab:traceStats}.

\subsubsection{End-to-End Delay Analysis}
\label{sec:e2edly}
\begin{figure}
	\subfigure[Low-quality Jurassic Park 1.]{
		\label{fig:e2eLowDly1}
		\includegraphics[scale=0.60]{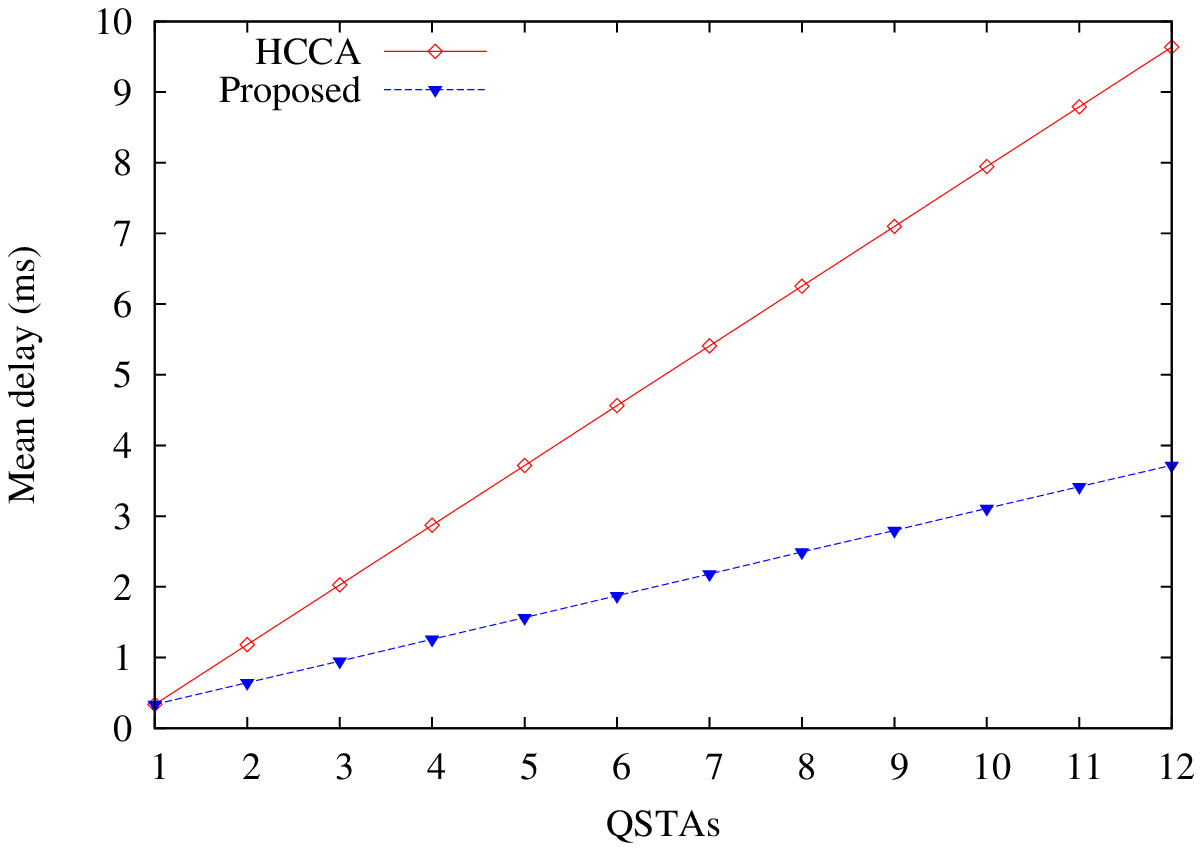}
	}
	\subfigure[High-quality Jurassic Park 1.]{
		\label{fig:e2eHighDly1}
		\includegraphics[scale=0.60]{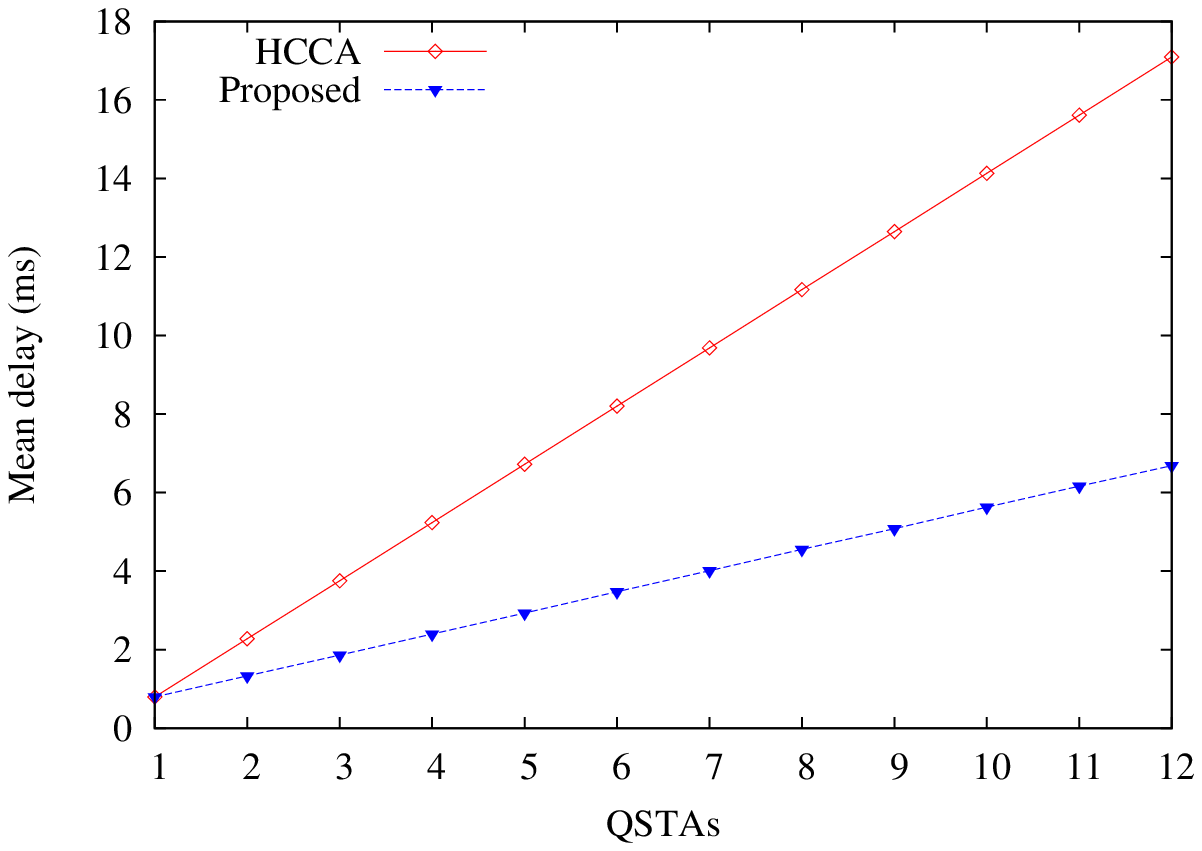}
	}
	\subfigure[Low-quality Formula 1.]{
		\label{fig:e2eLowDly2}
		\includegraphics[scale=0.60]{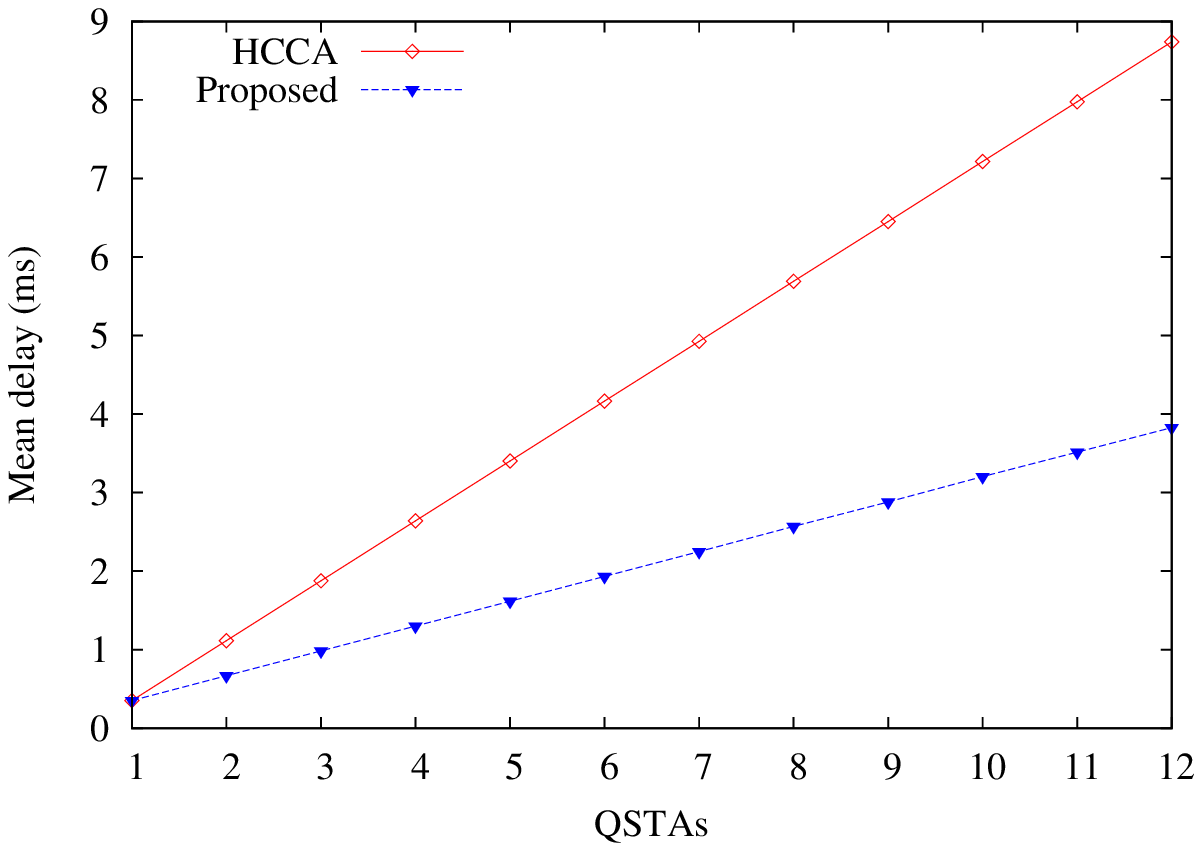}
	}
	\subfigure[High-quality Formula 1.]{
		\label{fig:e2eHighDly2}
		\includegraphics[scale=0.60]{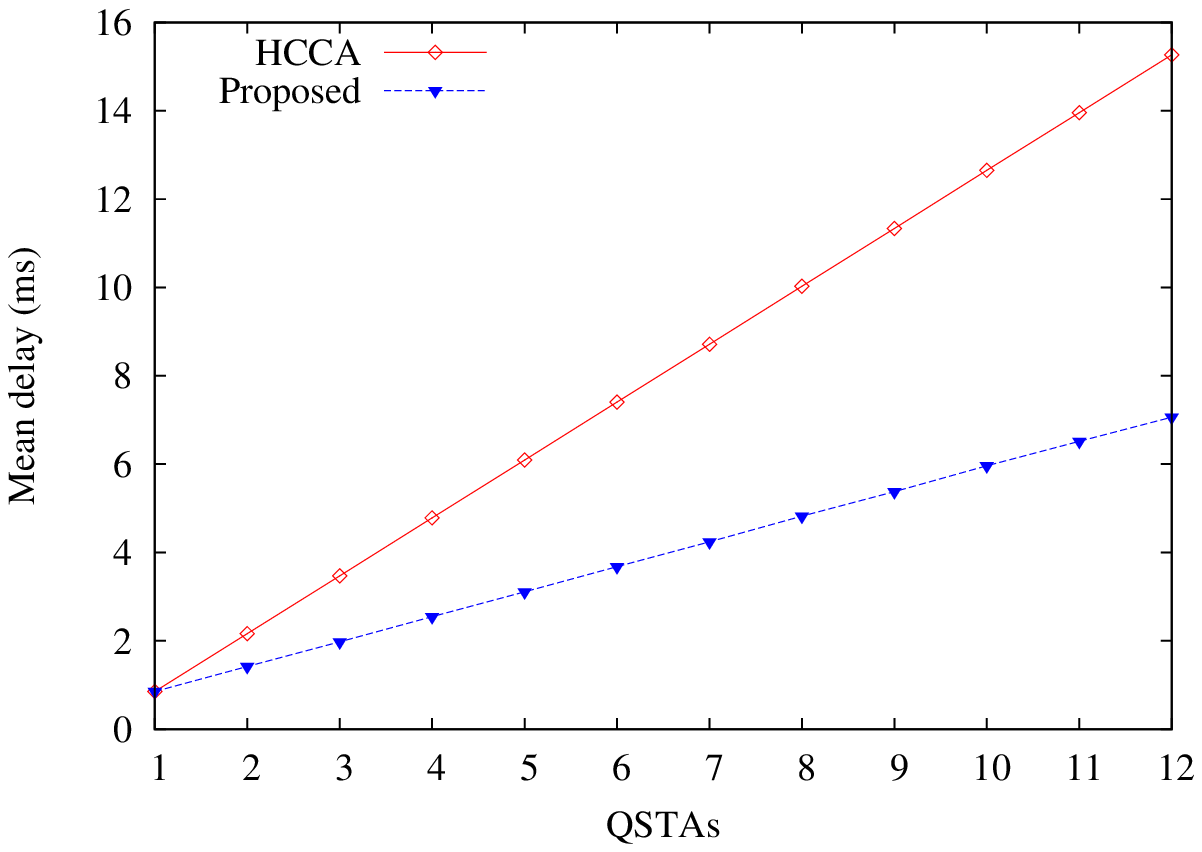}
	}
	\caption{Mean end-to-end delay as a function of the number of stations.}
\end{figure}

The end-to-end delay is defined as the time elapsed from the generation of the packet at the source \gls{QSTA} application layer until it has been received at the \gls{QAP} and is expressed in Equation~\eqref{eq:meanDelay}.
\begin{eqnarray}
	\label{eq:meanDelay}
	e2eDelay=\frac { \sum_{i=1}^{N} ( R_{i}-G_{i})  } {N},
\end{eqnarray}
where $G_{i}$ is the generation time of packet $i$ at the source \gls{QSTA}, $R_{i}$  is the receiving time of the particular packet at the \gls{MAC} layer of the \gls{QAP}, and $N$ is the total number of packets for all flows in the system. The end-to-end delay has been measured for the three video types to study the efficiency of  both \gls{HCCA} and our schedulers with different traffic variability. Figure~\ref{fig:e2eLowDly1}, \ref{fig:e2eHighDly1} and \ref{fig:e2eLowDly2}, \ref{fig:e2eHighDly2} depict the delay experienced by data packets for the low, medium and high quality video, respectively. One can notice that the end-to-end delay boosts with the increase of the packet size, higher quality exhibit higher end-to-end delay and vice versa. The increase of the delay in higher quality videos can be justified by the large amount of the  allocated to each \textit{TS}, as in Equation~\eqref{eq:txop}, which leads to maximize the wasted \gls{TXOPs} that keep the subsequent \textit{TSs} awaiting in their transmission queue longer time. It is worth noting that the proposed scheduler achieved about 52\% and 46\% delay improvement over \gls{HCCA} for Jurassic Park 1 and Formula 1 respectively. The reason of achieving better improvement in Jurassic Park 1 is the higher packet size comparable to that in Formula 1 thereby the granted \gls{TXOP} obtained in \gls{HCCA} is far from the needed \gls{TXOP} which in turn causes higher packet delay. Furthermore, the delay improvement in our scheduler is justified by the accurate calculation of the \gls{TXOP}. Unlike the \gls{HCCA} scheduler that only relies on the mean traffic characteristic which is not reflecting the actual traffic behavior.

\subsubsection{Throughput Analysis}
\begin{figure}
	\subfigure[Low-quality Jurassic Park 1.]{
		\label{fig:thrpLow1}
		\includegraphics[scale=0.60]{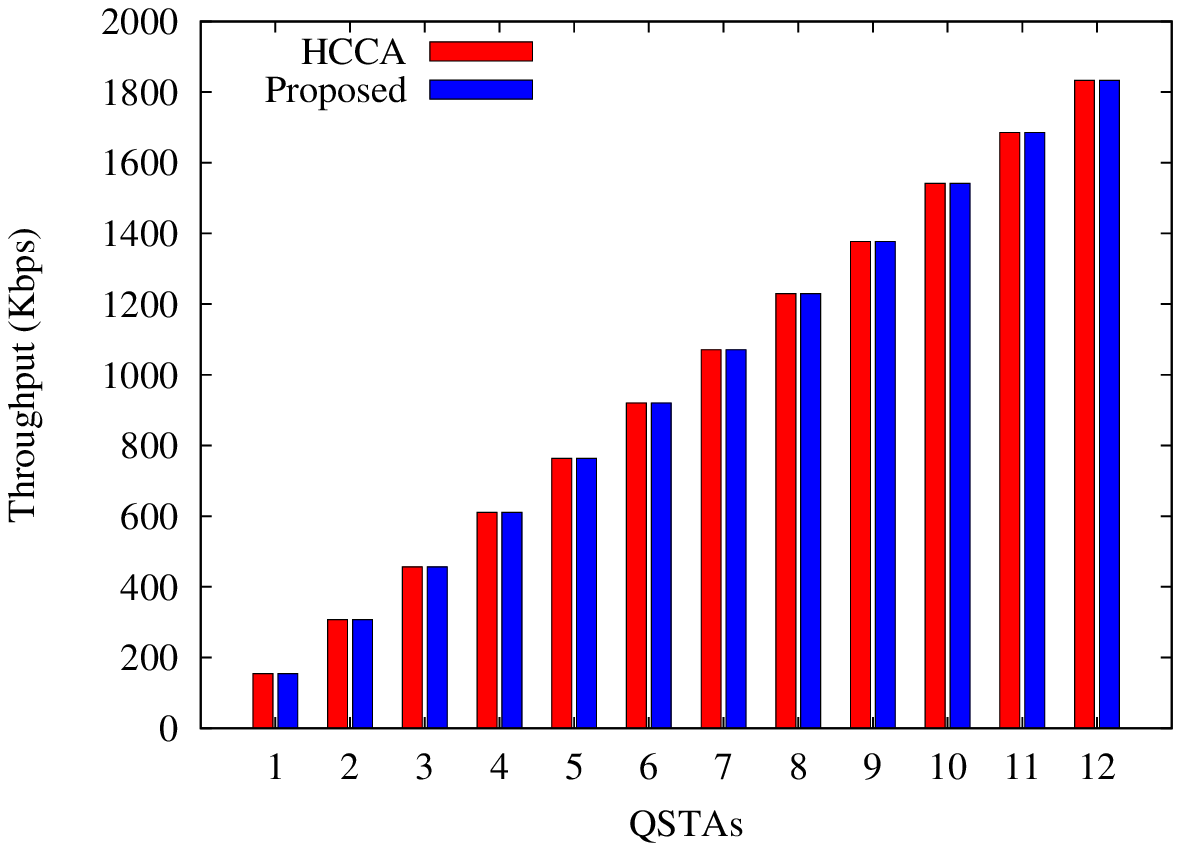}
	}
	\subfigure[High-quality Jurassic Park 1.]{
		\label{fig:thrpHigh1}
		\includegraphics[scale=0.60]{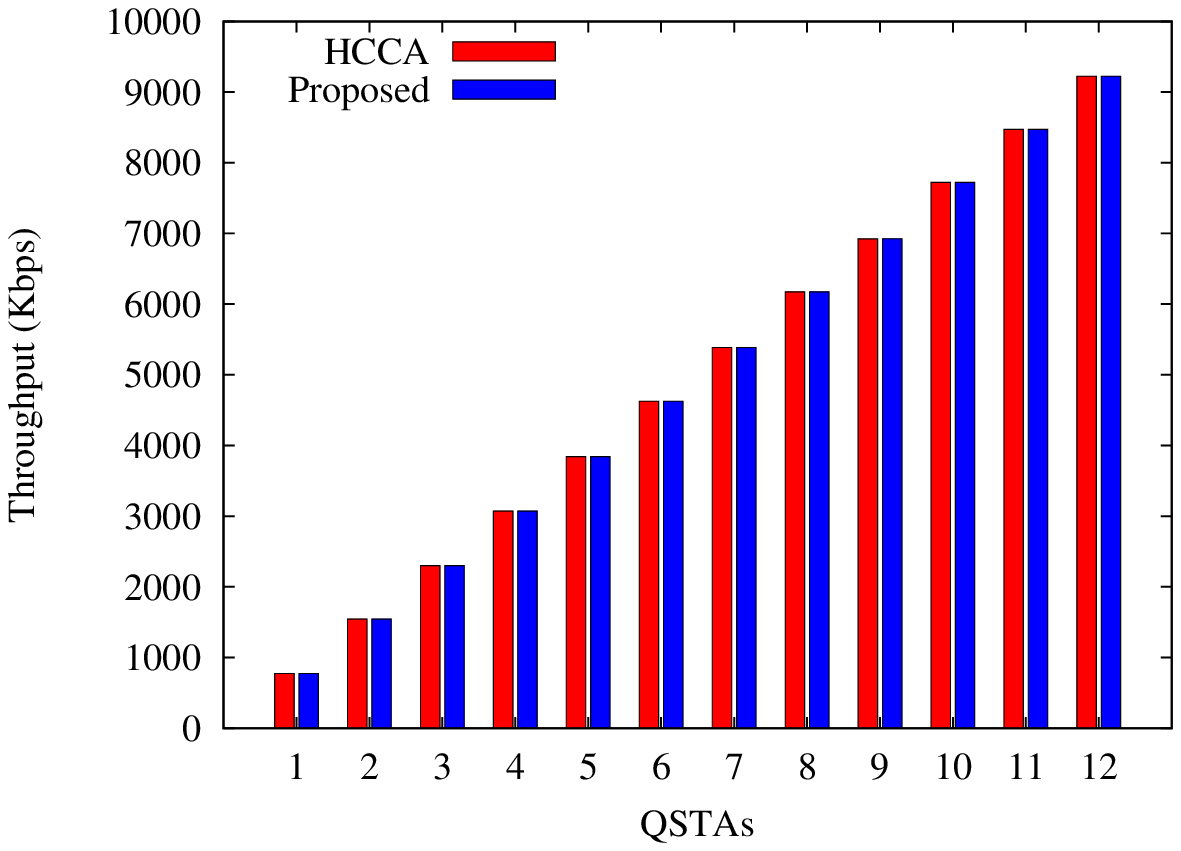}
	}
	\subfigure[Low-quality Formula 1.]{
		\label{fig:thrpLow2}
		\includegraphics[scale=0.60]{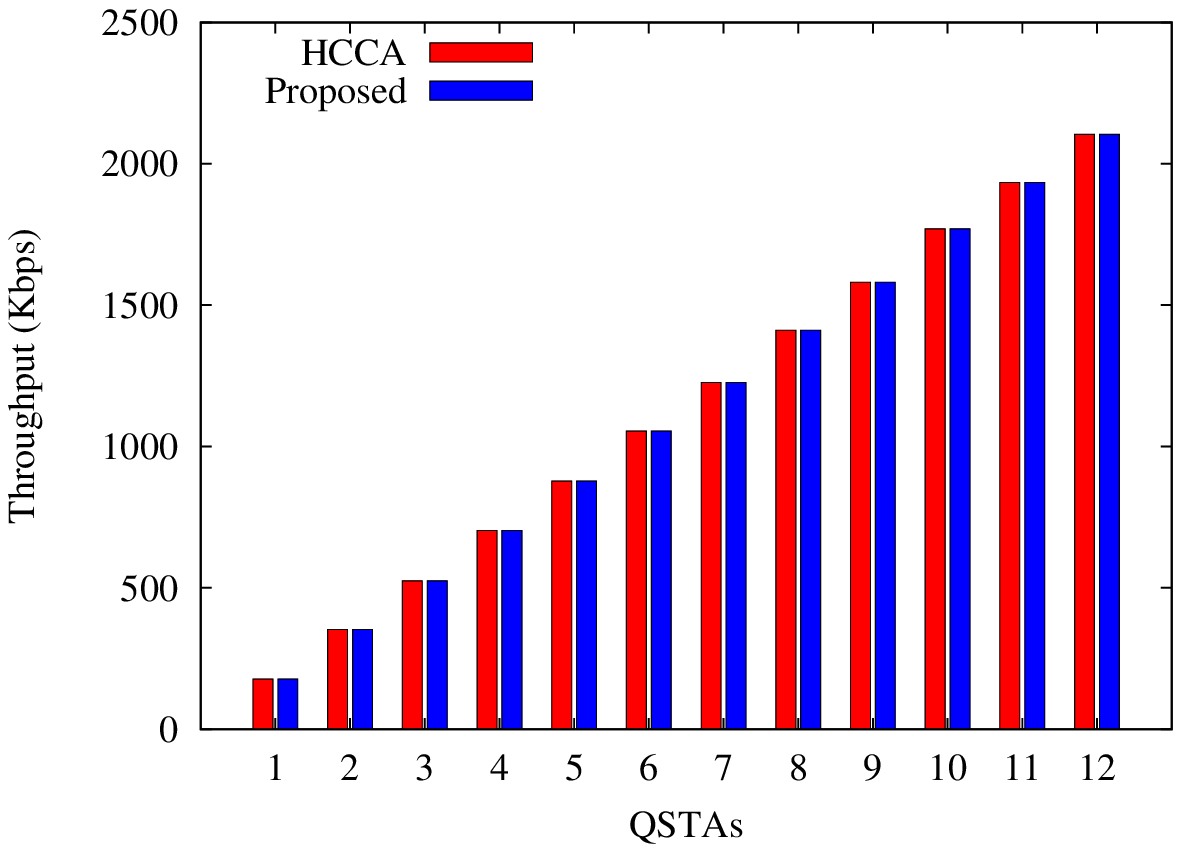}
	}
	\subfigure[High-quality Formula 1.]{
		\label{fig:thrpHigh2}
		\includegraphics[scale=0.60]{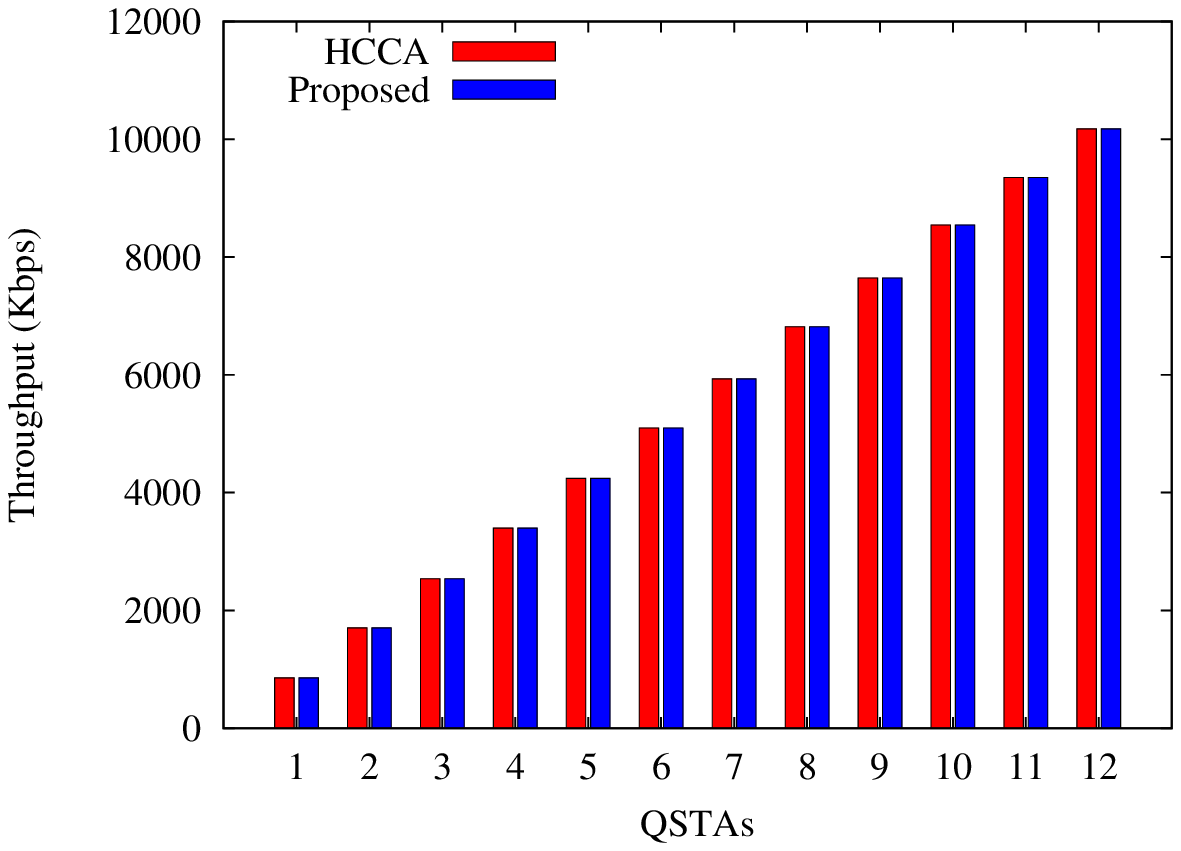}
	}
	\caption{Aggregate throughput as a function of the number of stations.}
\end{figure}

The aggregate throughput of the examined schedulers has been investigated against the number of stations. This is to verify that our scheduler is efficient in supporting \gls{QoS} for \gls{VBR} traffics which maintaining the utilization of the channel bandwidth. The aggregate throughput is calculated using Equation~\eqref{eq:throughput}.
\begin{eqnarray}
	\label{eq:throughput}
	AggregateThrp=\frac{\sum_{i=1}^{N}(Size_{i})} {time},
\end{eqnarray}
where $Size_{i}$ is the received packet size at the \gls{QAP}, $time$ is the simulation time and $N$ is the total number of the received packets at \gls{QAP} during the simulation time. Figure~\ref{fig:thrpLow1},  \ref{fig:thrpHigh1}, \ref{fig:thrpLow2} and \ref{fig:thrpHigh2} depict the aggregate throughput with increasing the network load for the low-, medium- and high-quality Jurassic Park 1 videos,  respectively. The results show that the throughput is the same as that achieved by the \gls{HCCA} scheduling scheduler. This implies that our approach enhanced the end-to-delay without jeopardizing the channel bandwidth.
\subsubsection{Aggregate TXOP Duration}
\begin{figure}
	\subfigure[Low-quality Jurassic Park 1.]{
		\label{fig:TXOPLow1}
		\includegraphics[scale=0.60]{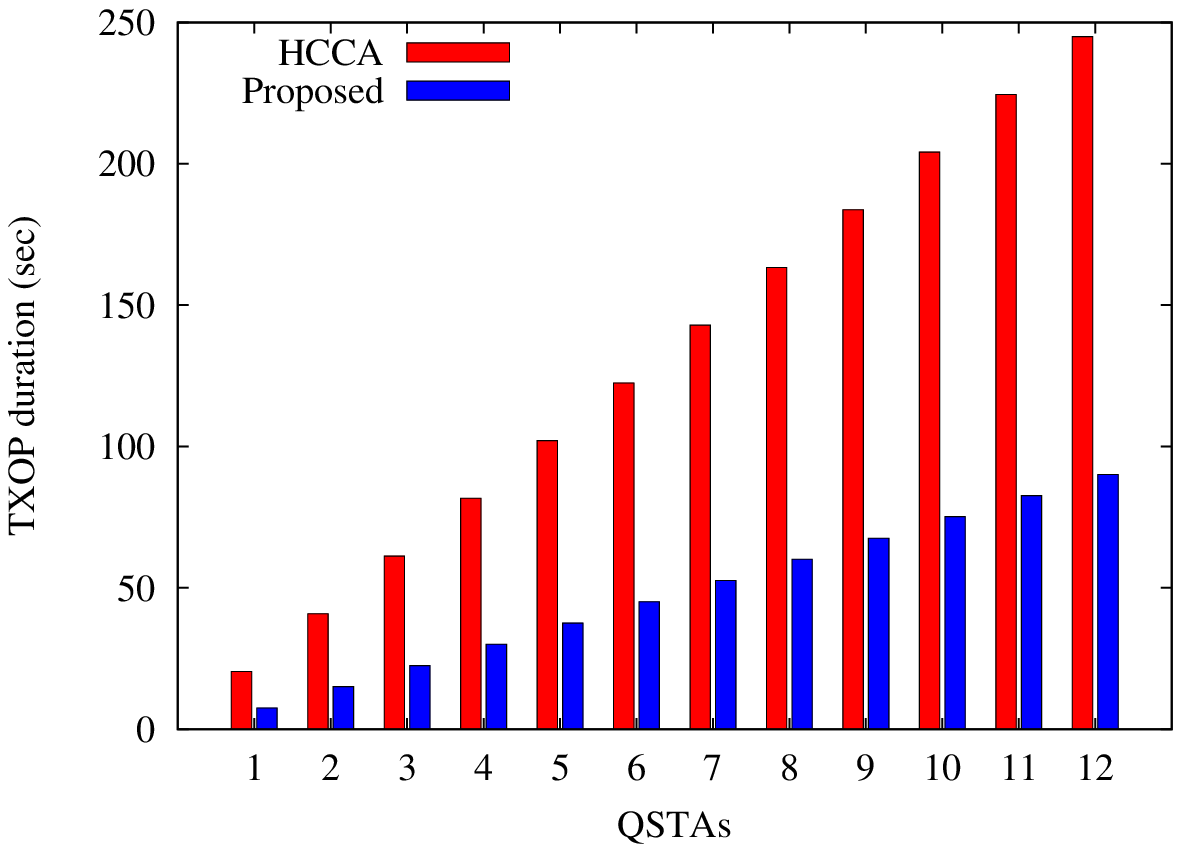}
	}
	\subfigure[High-quality Jurassic Park 1.]{
		\label{fig:TXOPHigh1}
		\includegraphics[scale=0.60]{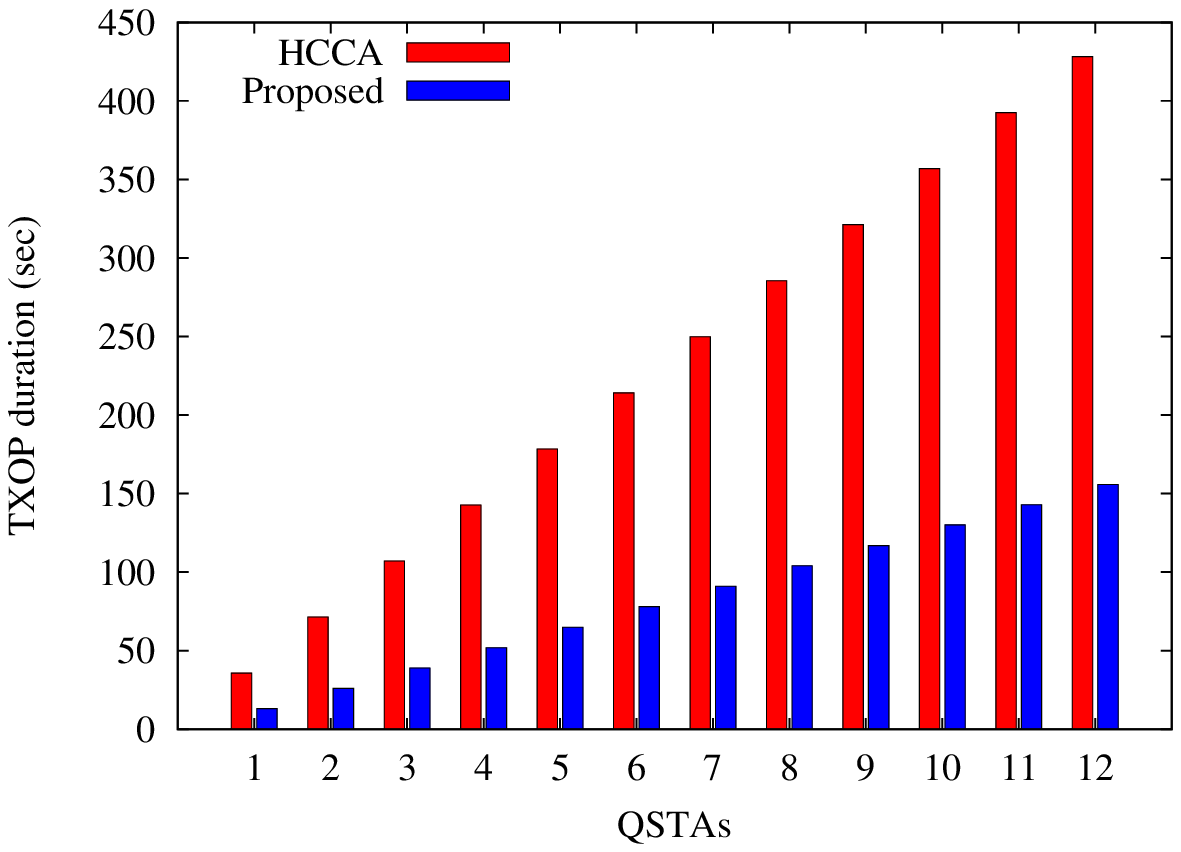}
	}
	\subfigure[Low-quality Formula 1.]{
		\label{fig:TXOPLow2}
		\includegraphics[scale=0.60]{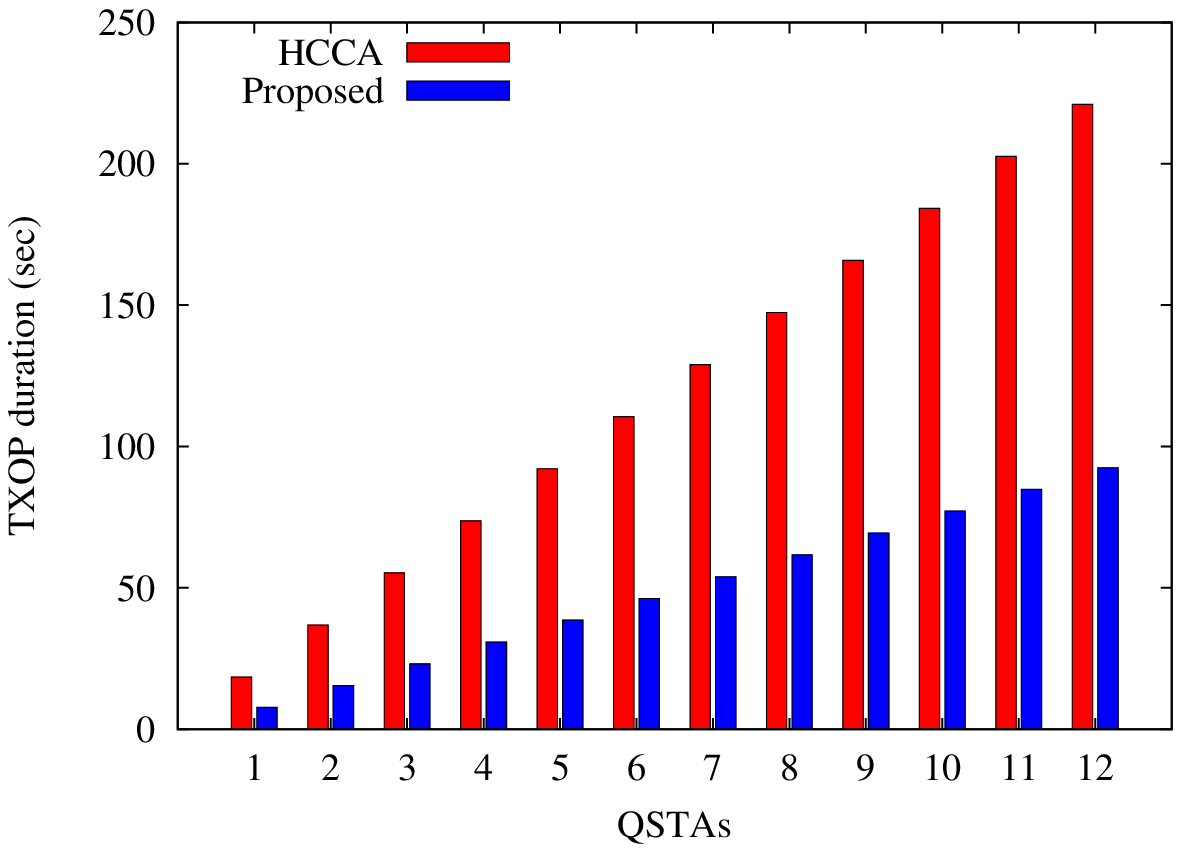}
	}
	\subfigure[High-quality Formula 1.]{
		\label{fig:TXOPHigh2}
		\includegraphics[scale=0.60]{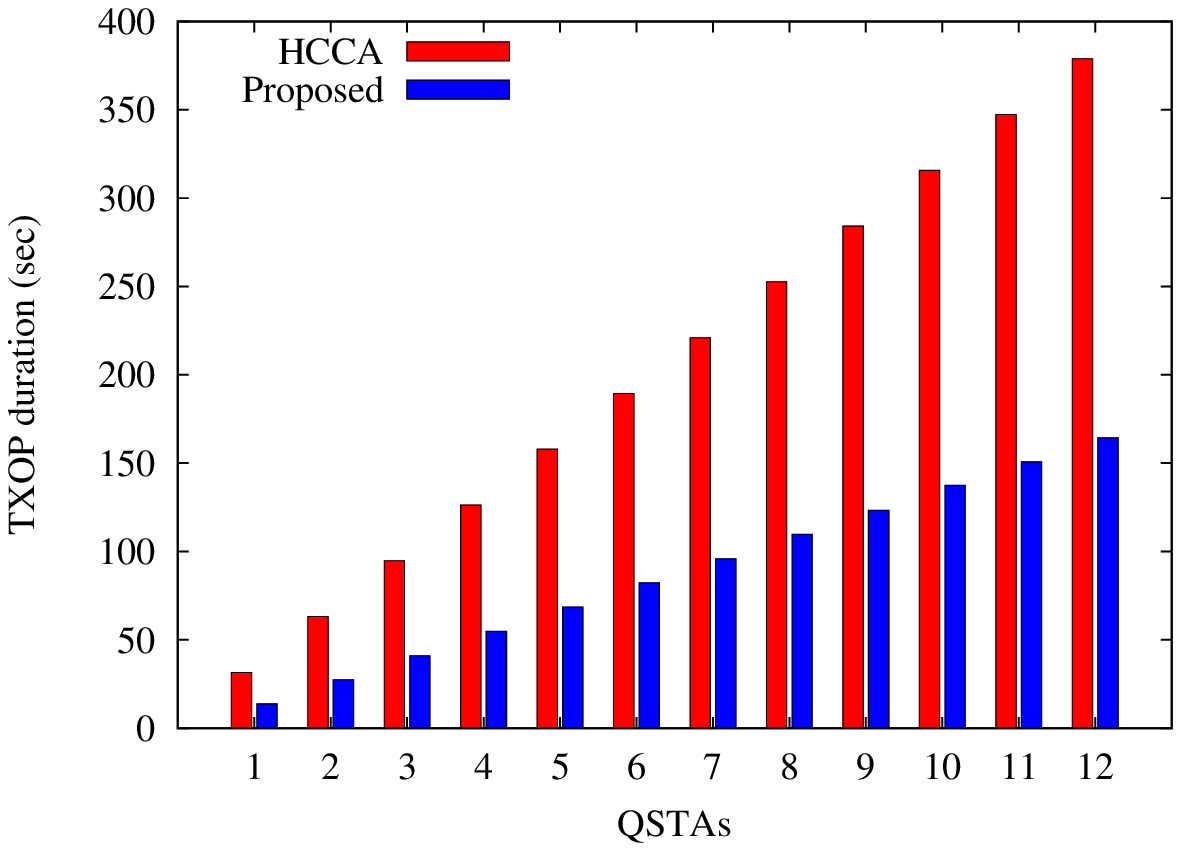}
	}
	\caption{Aggregate \gls{TXOP} duration as a function of the number of stations.}
	\label{fig:AggTXOP}
\end{figure}

To investigate the efficiency attained by the proposed scheduler in supporting prerecorded videos against \gls{HCCA}, the aggregate \gls{TXOP} duration is measured and can be defined as the total of \gls{TXOP} duration granted to all \glspl{QSTA} for the simulation time in units of seconds. In Figure~\ref{fig:AggTXOP}, the aggregate \gls{TXOP} is shown in the examined videos with increasing the network load. For Low-quality videos Figures~\ref{fig:TXOPLow1} and \ref{fig:TXOPLow2} demonstrate that allocating fixed \gls{TXOP} for all \textit{TS} frames in \gls{HCCA} might exceed the need of the traffic. In that case only a small portion of the granted \gls{TXOP} is exploited resulting in what's called wasted \gls{TXOPs}. On the contrary, the proposed scheduler operates according to the  actual information about frame size, the granted \gls{TXOP} is considerably smaller than that in \gls{HCCA} without jeopardizing the throughput. This fact is more obvious when transmitting High-quality videos (Figures~\ref{fig:TXOPHigh1} and \ref{fig:TXOPHigh2}) where the wasted \gls{TXOP} is much higher. It is worth mentioning that Jurassic Park 1 shows higher wasted \gls{TXOP} than that in Formula 1 as the CoV is higher.

\section{Conclusion}
\label{sec:conclusion}
This study proposed a novel scheduling scheduler to support \gls{VBR} video streams in IEEE 802.11e WLANs. This scheduler dynamically assigns \gls{TXOP} to a \gls{QSTA} based on piggybacked information about the next frame size with each packet sent of uplink traffic instead of assigning fixed \gls{TXOP} of \gls{HCCA}. Accordingly, \gls{HC} is able to poll \glspl{QSTA} with regard to their fast changing in the traffic profile  so as to prevent \glspl{QSTA} from receiving unnecessary large \gls{TXOP} which produces a remarkable increase in the packet delay. The proposed scheduler has been evaluated over two video streams with varying quality level to verify the performance of supporting videos with low and high variability traffics. Simulation results reveal the efficiency of the proposed scheduler over \gls{HCCA} in terms of minimizing the end-to-end delay while maintaining the system throughput and enhance the channel bandwidth utilization as well.

\section*{Acknowledgements}

This research work was supported by the Malaysian Ministry of Education under the Fundamental Research Grant Scheme, FRGS/1/2014/ICT03/UPM/01/1
\bibliographystyle{unsrt}
\bibliography{references}      

\end{document}